\documentclass[hyper,11pt,letterpaper]{JHEP3}

\usepackage{amsmath,amssymb,epsfig}
\newcommand{\bc}{\begin{center}}
\newcommand{\ec}{\end{center}}
\newcommand{\ba}{\begin{array}}
\newcommand{\ea}{\end{array}}
\newcommand{\beq}{\begin{equation}}
\newcommand{\eeq}{\end{equation}}
\newcommand{\bea}{\begin{eqnarray}}
\newcommand{\eea}{\end{eqnarray}}
\newcommand{\bmx}{\begin{pmatrix}}
\newcommand{\emx}{\end{pmatrix}}
\newcommand{\nn}{\nonumber}

\newcommand{\del}{\partial}
\newcommand{\half}{\frac{1}{2}}

\newcommand{\eref}[1]{Eq.~(\ref{#1})}

\newcommand{\cP}{{\cal P}}

\newcommand{\zbar}{{\bar z}}
\newcommand{\wbar}{{\bar w}}
\newcommand{\Wbar}{{\bar W}}
\newcommand{\cbar}{{\bar c}}
\newcommand{\cT}{{\cal T}}

\newcommand{\tX}{{\tilde X}}

\def\IB{\relax{\rm I\kern-.18em B}}
\def\IC{{\relax\hbox{\kern.3em{\cmss I}$\kern-.4em{\rm C}$}}}
\def\ID{\relax{\rm I\kern-.18em D}}
\def\IE{\relax{\rm I\kern-.18em E}}
\def\IF{\relax{\rm I\kern-.18em F}}
\def\II{\relax{\rm I\kern-.18em I}}
\def\Id{\relax{1\kern-.32em 1}}
\def\IG{\relax\hbox{$\inbar\kern-.3em{\rm G}$}}
\def\IR{\relax{\rm I\kern-.18em R}}
\newcommand\sfrac[2]{{\textstyle\frac{#1}{#2}}}

\normalsize

\newcommand{\bm}{\bar{m}}
\def\ga{\gamma}
\def\Ga{\Gamma}

\title{FZZ Algebra}
\preprint{hep-th/0606037\\ TIFR/TH/06-12 \\YITP-SB-06-10}

\author{Anindya Mukherjee\footnote{Email:
anindya\_m@theory.tifr.res.in}$^{~1}$,
Sunil Mukhi\footnote{Email: mukhi@tifr.res.in}$^{~1}$ and Ari
Pakman\footnote{Email: ari.pakman@stonybrook.edu}$^{~2}$\\
\it $^1$Tata Institute of Fundamental Research,\\
\it Homi Bhabha Rd, Mumbai 400 005, India\\

\it $^2$C.N. Yang Institute for Theoretical Physics,\\
\it Stony Brook University, \\
\it Stony Brook, NY 11794-3840, USA}
\abstract{The duality between the Sine-Liouville conformal field
theory and the two dimensional black hole is revisited by considering
the two possible Sine-Liouville dressings together. We show that this
choice is consistent with the structure of correlation functions, and
that the OPE of the two dressings yields the black hole deformation
operator. As an application of this approach, we investigate the role
of higher winding perturbations in the context of $c=1$
strings, where we argue that they are related to higher-spin discrete
states that generalize the 2d black hole operator.}


\begin{document}

\section{Introduction}
It has long been known that the bosonic string admits a
two-dimensional black-hole like background, described as a
gauged $SL(2,R)/U(1)$ WZW model\cite{Witten:1991yr} and can also be
thought of for some values of the parameters as a solution of the
lowest order (in $\alpha'$) effective action\cite{Mandal:1991tz,
Elitzur:1991cb}. Moreover, it was shown that when viewed as a
perturbation of the $c=1$ string theory, the leading term in this solution
uniquely extends to a full solution of closed string field
theory\cite{Mukherji:1991kz}.

Some years ago, Fateev, Zamolodchikov and
Zamolodchikov\cite{FZZ} proposed that the gauged $SL(2,R)/U(1)$ CFT has a
dual description in terms of a free theory (with a linear dilaton)
perturbed by a Sine-Liouville potential. This remarkable relation
between two seemingly different models, the so called FZZ duality, has
been explored and applied in several ways (see for example
\cite{Kazakov:2000pm, Fukuda:2001jd, Giveon:2001up,Giribet:2004zd,
Hikida:2004mp, Bergman:2005qf}).  The duality has an $N=2$
supersymmetric version~\cite{Giveon:1999px, Hori:2001ax, Tong:2003ik},
as well as a realization on the boundary of the
worldsheet~\cite{Israel:2004jt,Lukyanov:2005bf}.

On the other hand, progress made during the last few years in the
study of non-rational conformal field theories (see
\cite{Teschner:2001rv,Nakayama:2004vk,Schomerus:2005aq} for reviews)
has shown that both dressings of Liouville-like perturbations in
linear dilaton theories appear in the exact solutions
\cite{Zamolodchikov:1995aa}. The latter typically have two classical
limits, and in each limit one of the two perturbations disappears.
This suggests that the classically vanishing operator is a
non-perturbative quantum effect generated by the backreaction of the
first one.

Therefore it is natural to consider a Sine-Liouville theory where both
dressings are taken into account and to ask how the FZZ duality fits
in such a setting.  In this work we propose an answer to this question
which gives a new perspective on the FZZ duality.  Our approach is
based on the observation that the OPE of the two Sine-Liouville
dressings yields the black hole perturbation. Therefore the latter
operator closes a sort of algebra which we have dubbed the FZZ algebra.
In order to preserve the exact marginality of the
perturbations, the black hole operator should then be added to the action.
Finally, a perturbative computation will show that the coefficient
in front of the second sine-Liouville should be put to zero.
In this way, the standard form of the FZZ duality is recovered,
with just one Sine-Liouville perturbation along with that of the black hole.

This approach to the FZZ duality suggests in turn a natural
generalization of the FZZ algebra in the $c=1$ non-critical string
context. This is motivated by the fact that in this case, the
two-dimensional black hole is the first of an infinite family of
solutions to the closed string field theory equations, each one
corresponding to one of the discrete states of the $c=1$
string~\cite{Mukherji:1991kz}.  We find that when Sine-Liouville
perturbations with different winding numbers are turned on, all the
discrete states of the $c=1$ can be generated by multiple OPEs.  This
strongly points to the existence of an infinitely generalized FZZ
duality in the $c=1$ string, which should be further investigated.

The organization of this work is as follows. In Section 2 we briefly
review the two-dimensional black hole background and the FZZ duality.
In Section 3 we introduce the FZZ algebra, we show that all the
interactions involved are compatible with the parafermionic symmetry
of the $SL(2,R)/U(1)$ coset and that the second Sine-Liouville
dressing is consistent with the correlation functions of the
theory. In Section 4, we briefly review the $c=1$ string theory and
present our proposal for the enlargement of the FZZ algebra in this
model.  Section 5 contains the conclusions.

\section{Euclidean 2d Black Hole and FZZ Duality}

\subsection{2d Black Hole - Review}

We start by reviewing some basic properties of the two dimensional
cigar or black hole solution in noncritical string theory
\cite{Witten:1991yr,Mandal:1991tz,Elitzur:1991cb}
This will also serve to establish some notation and conventions.

This black hole solution can be written as an exact conformal field
theory (all orders in $\alpha'$), namely an $SL(2,R)/U(1)$ WZW model
\cite{Witten:1991yr}, whose Euclidean version is a $\sigma$-model with
metric:
\bea
\nonumber
ds^2 &=& k\left( (1-e^{2Q\phi})\,dt^2 +
\frac{1}{1-e^{2Q\phi}}d\phi^2\right) \,,
\\
\Phi-\Phi_0 &=& Q\phi,\qquad -\infty <\phi < 0 \,\,.
\eea
Here $k$ is the level of the $SL(2,R)$ WZW model and
$Q=\frac{1}{\sqrt{k-2}}$.

By a change of coordinates, this solution can also be written:
\begin{eqnarray}
\nonumber
ds^2 &=& k\left(dr^2 + \tanh^2 r\,d\theta^2\right)\\
\Phi-\Phi_0 &=& -2\log \cosh r,\qquad -\infty < r < 0
\,.
\end{eqnarray}
The geometry of the Euclidean black hole is that of a cigar ending at
$r=0$. Its asymptotic radius as $r\to -\infty$ is
\beq
R=\sqrt{k} \,.
\eeq
The value of the dilaton at the tip, $\Phi_0$, can be identified with
the mass of the black hole:
\beq
M \sim e^{-2\Phi_0} \,.
\eeq
The 2d black hole can either be considered by itself as a string
background, or adjoined to another ``internal'' CFT to form the total
string background. In the former case, conformal invariance of the
worldsheet theory requires:
\beq
c_{tot} = \frac{3k}{k-2}-1=26 ~~\Rightarrow~~
k=\sfrac94~~\Rightarrow~~ R=\sfrac32 \,.
\label{ccharge}
\eeq
Therefore in this case one is in the regime of small $k$, and the
spacetime solution is not very reliable. On the other hand, if we add
an internal CFT then it is easy to see that $k$ can be arbitrarily large
and one expects the spacetime solution to be a reliable guide to the
physics. In this and the next few sections we will assume the most general
situation, with $k$ arbitrary. Later we will specialise to the case
where there is only a black hole and no internal CFT.

An important role will be played by the fact that the 2d black hole
background has a parafermionic $SL(2,R)/U(1)$ symmetry.  Note that a
cosmological Liouville perturbation would spoil this symmetry.  Hence
we assume there is no cosmological perturbation, which is physically
acceptable since the would-be strong coupling region is already cut
off by the black hole geometry.

For large negative $\phi$, the black hole metric can be written
\bea
\label{bhpert}
ds^2 &=& k\Big( (1-e^{2Q\phi})\,dt^2 + (1+e^{2Q\phi}) d\phi^2\Big)\,,\nn\\
&=& k\Big(dt^2 + d\phi^2 - (dt^2 - d\phi^2)e^{2Q\phi}\Big)\,.
\eea
Thus, infinitesimally the black hole is generated by a perturbation
\bea
\Delta S = (\del X{\bar\del}X - \del\phi{\bar \del}\phi)e^{2Q\phi} \,.
\eea
The second term is a pure gauge in BRST cohomology. Therefore the
black hole background is generated by the operator:
\beq
B =  \del X{\bar\del}X e^{2Q\phi} \,.
\eeq
It should be kept in mind that this operator only describes the 2d
black hole far away from the horizon, in the weak-coupling region
$\phi\to-\infty$. However, it unambiguously generates the full
solution, in the sense that a CFT perturbed by $B$ will flow to the
CFT of the Euclidean 2d black hole\cite{Mukherji:1991kz}. In this
process the spacetime gets cut off at the horizon, leading to the
well-known property that winding number is violated: a string wrapped
around the Euclidean time direction in the asymptotic region can be
slipped off at the horizon. Violation of winding number is not,
however, evident from inspection of the operator $B$, which by itself
conserves winding number.

\subsection{FZZ Duality}

The FZZ duality \cite{FZZ} states that the Euclidean 2d black hole
discussed above is ``dual'' to the Sine-Liouville perturbation of the
linear dilaton theory.

The latter arises by coupling a compact ``matter'' coordinate $X$ to
the Liouville field. Since $X$ is compact, it can be split into $X=X_L
+ X_R$.  Let us normalize the holomorphic fields $X_L$ and $\phi(z)$
as~($\alpha'=1$)
\bea
X(z) X(w) \sim \phi(z) \phi(w) \sim  -\frac12 \log (z-w) \,.
\label{normalization}
\eea
and similarly for the anti-holomorphic fields $X_R, \phi(\bar{z})$.
The worldsheet stress tensor is
\bea
T = -(\del X)^2  - (\del \phi )^2  + Q \del^2 \phi,
\eea
with
\bea
Q = \frac{1}{\sqrt{k-2}} \,,
\eea
and the central charge is
\beq
c= 2 + 6Q^2 \,,
\eeq
which is the same as \eref{ccharge}. The linear dilaton is
given by
\beq
\Phi-\Phi_0 = Q\phi \,,
\eeq
so that, with $g_s=e^{\Phi_0}e^{Q\phi}$, the theory is weakly coupled
as $\phi\to-\infty$.

The vertex operators of this theory are written:
\beq
V_{\alpha,\beta} = e^{2i\alpha X}e^{2\beta\phi}\,,
\eeq
and have conformal dimension
\beq
\Delta= \alpha^2 + \beta(Q-\beta) \,.
\eeq
The wave function corresponding to these operators is obtained by
multiplying by $g_s^{-1}\sim e^{-Q\phi}$. It follows that whenever
$\beta<\frac{Q}{2}$ the wave function is non-normalizable, in that it
is peaked about the weak-coupling region $\phi\to-\infty$. This is
sometimes called the ``allowed'' dressing. Its insertion creates a
local deformation of the worldsheet.  For $\beta>\frac{Q}{2}$ the wave
function decays at weak coupling and is normalizable, and its
insertion creates a non-local deformation.

If the theory is perturbed by an operator that creates a ``wall'' at
strong coupling, the situation is different. Only one linear
combination of right-moving and left-moving waves survives.  As a
consequence, the corresponding Euclidean operator will be a linear
combination of normalizable and non-normalizable ones.

Now let us introduce the Sine-Liouville perturbations:
\beq
\cT_{\pm R}^+  = e^{\pm iR(X_L-X_R)}\,e^{(Q-|Q-\frac{1}{Q}|)\phi} \,,
\label{slpertplus}
\eeq
where as before, $R=\sqrt{k}$. The subscript labels the ``winding
momentum'' for the matter part of the vertex operator, while the sign
in the superscript labels the Liouville dressing. In
particular, the above operators both have the ``allowed'' value of the
Liouville dressing, so that the corresponding wave-functions grow at
weak coupling and are non-normalizable. These operators carry winding
number $\pm1$ around the Euclidean time direction.

%Conventionally it is only the non-normalisable Sine-Liouville operator
%that is used to perturb the linear dilaton theory. Thus, as usually
%stated, the Sine-Liouville theory is the linear dilaton background
%perturbed by $\lambda \cT_{R}^+ + {\bar\lambda} \cT_{-R}^+$.

The FZZ duality states that the 2d black hole theory is
equivalent to Sine-Liouville. One of our goals in what follows will be
to make this notion more precise. However first let us review the
existing evidence for this duality. It comes from the knowledge of the
exact two- and three-point functions (on the sphere) of the 2d black
hole theory. For example, the two-point function, which we will
re-obtain below, is:
\begin{eqnarray}
\label{twopointf}
R(j,m,\bm)
= \left( \frac{ \mu \pi\Ga(\frac{1}{k-2})}{\Ga(1-\frac{1}{k-2})}
\right)^{1-2j}
\frac{\Gamma(2j-1)\Gamma(1+\frac{2j-1}{k-2})} {\Gamma(-2j+1)
\Gamma(1-\frac{2j-1}{k-2})}
\frac{\Gamma(-j+1+ \bm)\Gamma(-j+1-m)}{\Gamma(j+\bm)
\Gamma(j-m)} \,.
\nn \\
\end{eqnarray}
The poles in the first two $\Gamma$-functions of the numerator reflect
the noncompact nature of the target space \cite{Goulian:1990qr}. It
can be shown that the positions of the poles of the first
$\Gamma$-function, occurring at
\beq
j=0, -\sfrac12,-1,-\sfrac32,\cdots
\eeq
can be obtained using the black hole operator as a screening charge,
and the residues at these poles can be computed using free field
techniques. Together this determines the above correlator.
On the other hand, the poles of the second $\Gamma$-function, at
\beq
1+ \frac{2j-1}{k-2} = 0,-1,-2,\cdots
\eeq
can be obtained using the Sine-Liouville operator as the screening
charge, and their residues again give the remaining factors in the
correlator. The agreement has also been shown to hold
for three point functions in \cite{Fukuda:2001jd,Giribet:2004zd},
for processes conserving and violating winding number.

It is intriguing that this duality works quite similarly to channel
duality in critical string theory, where summing over the residues at
the $s$-channel poles gives the same answer as summing over the
residues at the $t$-channel poles. We are not aware if this similarity
has any further implications.

\section{FZZ Algebra}

Let us now study the linear dilaton theory with a Sine-Liouville
perturbation. In previous treatments it has been standard to add to
the worldsheet action just one of the two ``dressings'' of the
Sine-Liouville operator\footnote{In fact, in \cite{Kazakov:2000pm},
the dressing is chosen to connect to the semiclassical limit valid for
$Q\to\infty$. This is normalizable for $Q<1$ and becomes
non-normalizable for $Q>1$.}. Here we will start by considering
simultaneously both sine-Liouville dressings.
This point of view  has already
been demonstrated to be useful in similar contexts (see for example
\cite{Zamolodchikov:1995aa}).

As we will see shortly, this approach provides a direct link between
Sine-Liouville theory and the 2d black hole. We will demonstrate that
the linear dilaton theory perturbed by both dressings of
Sine-Liouville operators {\it requires} the black hole perturbation
operator to be turned on for consistency, i.e. exact marginality of
the perturbation. So it would seem that the true perturbed theory has both
Sine-Liouville and black hole operators turned on at the same time.
But this is not the end of the story.
A perturbative quantum computation will show
that the relative coefficients between
all the perturbations  are determined self-consistently,
in such a way that the coefficient of the second Sine-Liouville
perturbation should be set to zero.
In other words, the second Sine-Liouville
perturbation disappears after fulfilling the role
of forcing the black hole perturbation to be present.

We will first describe the general arguments justifying this
procedure, and will then show that using various different
combinations of the Sine-Liouville operators (of both dressings) and
black hole operator as screeners is consistent
with the structure of the correlators of the theory and fixes the
relative coefficients of the different perturbations.
A certain parafermionic symmetry will prove useful in
the discussion.

\subsection{The Algebra of the Interactions}

The perturbation to the action is as follows\footnote{This is
really Cosine-Liouville rather than Sine-Liouville.}
\beq
\label{slpert}
S\to S+ \int d^2z
\left(\cT_R^+ + \cT_{-R}^+ + \cT_R^- + \cT_{-R}^-\right)\,.
\eeq
Between them, these four terms incorporate both signs of the matter
momentum as well as both signs of the Liouville dressing. The
operators $\cT_{\pm R}^+$ were given in \eref{slpertplus}
while the other two are given by:
\beq
\label{slpertminus}
\cT_{\pm R}^- = e^{\pm iR(X_L-X_R)}\,e^{(Q+|Q-\frac{1}{Q}|)\phi} \,,
\eeq

As these operators are all of conformal dimension $(1,1)$, the above
perturbation is marginal to first order. Now consider the requirements
of exact marginality. The general rule is that the perturbations will
be exactly marginal if their OPE does not produce another $(1,1)$
operator\cite{Polchinski:1998rq}. However, if they do produce such an
operator then we are required to add that operator back into the
Lagrangian to restore marginality.

Now it is easily seen that the following OPE holds between
the mutually conjugate operators $\cT_R^+$ and $\cT_{-R}^-$ (a similar
relation holds between $\cT_{R}^-$ and $\cT_{-R}^+$):
\beq
\cT^+_{R}(z,\zbar)\cT^-_{-R}(w,\wbar)\sim
\frac{1}{|z-w|^2} \del X{\bar\del}X\,e^{2Q\phi} + \cdots
\eeq
Here we have exhibited only the $(1,1)$ operator appearing on the
RHS. More singular terms correspond to operators that are BRST
trivial. Even among the $(1,1)$ operators that can appear, we have
dropped BRST-trivial contributions such as
$\del\phi{\bar\del}\phi\,e^{2Q\phi}$.

On the right hand side of the above equation, we recognise the black
hole perturbation operator. This tells us that the Sine-Liouville
theory (when viewed as a perturbation of the original action by
operators of both Liouville dressings) is not by itself exactly
marginal, but marginality can be restored by including the
black hole perturbation. In turn, it is known that the latter
perturbation can be built up into a solution of closed string field
theory which, being unique, must be equivalent to Witten's
$SL(2,R)/U(1)$ CFT\footnote{This was demonstrated for $k=9/4 $ in
\cite{Mukherji:1991kz}.}.

It is worth noting that, as in \cite{Schmidhuber:1993qe}, the operators
generated by requiring exact marginality to second order are not quite
the physical operators, but rather some variants of them with an extra
multiplicative factor of the Liouville field $\phi$ in front. In the
present case the black-hole operator would be replaced by:
\beq
\del X \bar{\del} X e^{2Q\phi} \to \phi\, \del X \bar{\del} X e^{2Q\phi} \,.
\eeq
This is reminiscent of the fact that, at $c=1$, the cosmological
operator in the linear dilaton theory is not really $e^{Q\phi}$ but
$\phi\,e^{Q\phi}$. As in that case, the distinction between the
operator with and without a $\phi$ in front is expected to be
unimportant for a large class of explicit computations.

\subsection{Parafermionic symmetry}

To justify and spell out the above observations, we will now perform a
more explicit study of the linear dilaton theory perturbed by
Sine-Liouville and black hole operators.  A useful tool in this study
is the fact that when the dilaton slope of $\phi$ is
$Q=\frac{1}{\sqrt{k-2}}$ and the radius of the compact direction is
$R=\sqrt{k}$, the symmetry of the worldsheet is expanded from Virasoro
to the parafermionic $SL(2,R)/U(1)$ algebra.  A representation of this
symmetry in terms of the $\phi$ and $X$ bosons can be obtained by
adding one free boson $Z$, normalized as in (\ref{normalization}), and
starting with the following free-field representation of the level $k$
$SL(2,R)$ current algebra \bea
\label{sl}
J^3 &=& -\sqrt{k}\del Z \,, \\
J^{\pm}&=& (i\sqrt{k}\del X \mp \sqrt{k-2}\del \phi )\,
e^{\mp \frac{2}{\sqrt{k}}(iX-Z)}\,.
\nonumber
\eea
Since $J^3$ corresponds to the direction gauged to
obtain the  $SL(2,R)/U(1)$ coset, the parafermionic generators
can be obtained by dropping $Z$ from the above expressions.
This gives
\beq
\label{parafermions}
\psi^{\pm}= (i\sqrt{k}\del X \mp \sqrt{k-2}\del \phi )
e^{\mp \frac{2i}{\sqrt{k}}X} \,.
\eeq
The currents  (\ref{sl}) and (\ref{parafermions}) have similar
anti-holomorphic copies. A generic primary of the coset
can be written in terms of $SL(2,R)$ quantum numbers as
\bea
V_{j,m,\bm} = e^{2jQ \phi }
e^{- \frac{2i m}{\sqrt{k}}X_L}
e^{- \frac{2i\bm}{\sqrt{k}}X_R}\,,
\label{primary}
\eea
with
\bea
m= \frac{n+kw}{2} \qquad \bm = \frac{n-kw}{2}
\eea
where $n$ and $w$ are the momentum and the winding
of the $X$  direction.
This state has conformal dimensions
\bea
\Delta_{j,m} & =& -\frac{j(j-1)}{k-2} + \frac{m^2}{k} \,, \\
\bar{\Delta}_{j,\bm} & =&  -\frac{j(j-1)}{k-2} + \frac{\bm^2}{k}\,,
\eea
and descends to the $SL(2,R)/U(1)$ coset from an $SL(2,R)$ primary
with spin $j$ and $J^3_0 =m, \bar{J}^3_0=\bm$.

We are interested in turning on marginal perturbations
to the flat background which preserve the $SL(2,R)/U(1)$ symmetry.
Natural candidates are exponentials of $X$ and $\phi$.
Consider the OPE
\begin{eqnarray}
\psi^{\pm}(z) \, e^{iaX_L + b \phi}(w) &\sim& \frac{\mp b
\sqrt{k-2}}{(z-w)^{1 \pm \frac{a}{\sqrt{k}} }}
e^{b \phi(w) +i(a \mp \frac{2}{\sqrt{k}})X_L(w)}
\left(1 + O(z-w) \right) \nn \\
&\mp & \frac{k}{2} \del_z
\left(\frac{e^{\mp i \frac{2}{\sqrt{k}} X_L(z)+ iaX_L(w)}}
{(z-w)^{\pm \frac{a}{\sqrt{k}}}} \right)e^{b \phi(w)} \,.
\end{eqnarray}
Requiring mutual locality and no single pole fixes $a=w\sqrt{k}$,
with $w$ a non-zero
integer. Thus the perturbation will be a winding mode belonging to the
spectrum of the theory,
if  we combine left- and right-movers with opposite signs for $w$.
For each $w$, there are two values of $b$ which give an operator with
 $\Delta=\bar{\Delta}=1$. For the case of one unit of winding, the
Sine-Liouville operators are
\bea
S_{\pm}^1 \equiv \lambda_1\cT_{\pm R}^+
&=& \lambda_1 e^{\pm i \sqrt{k}(X_L-X_R) + \frac{1}{Q} \phi}
\label{sl1} \,,\\
S_{\pm}^{2} \equiv \lambda_2\cT_{\pm R}^-
&=& \lambda_2 e^{\pm i \sqrt{k}(X_L-X_R) + (2Q-\frac{1}{Q}) \phi } \,.
\label{sl2}
\eea
Turning on Liouville-like perturbations
in linear dilaton theories has the effect of screening the strong coupling region
$\phi \rightarrow +\infty$. This is indeed the case for $S_{\pm}^1$. For $S_{\pm}^2$,
this happens only when $2Q > 1/Q$, i.e., $k<4$. This region includes the $k=9/4$ value
corresponding to the pure two-dimensional black hole. Therefore,
we will trust the Lagrangian description
of the theory perturbed with $S_{\pm}^2$ in the region $k<4$, and resort to
the analytical continuation of the results otherwise.

The important point is that {\it both} operators are compatible with
the $SL(2,R)/U(1)$ symmetry.
%we conclude that
%the operators
%\bea
%e^{iw\sqrt{k}(X_L-X_R)}e^{2j_w^{\pm}Q \phi}
%\label{wind-pert}
%\eea
%with
%\bea
%j^{\pm}_{w}= \frac12 \pm  \sqrt{1 + \frac{w^2k -4}{Q^2}}\,,
%\eea
%can be added to the flat background
%while preserving the $SL(2,R)/U(1)$ symmetry.
%The operators with $j^+_w$ are normalisable and those with
%$j^-_w$ are non-normalisable.
Now, the chiral black hole perturbation $\del X e^{2 Q \phi}$ should
also be compatible with the $SL(2,R)/U(1)$ symmetry. This is indeed
the case, but happens only at a fixed point of the gauge orbit of the
BRST trivial state $\del \phi e^{2 Q \phi} $.  To find this point,
consider
\bea
B = \left( \del X + \alpha \del \phi \right)  e^{2 Q \phi} \,.
\eea
Its OPE with $\psi^+$ is
\bea
\psi^+(z) B(w) \sim e^{-2i \frac{X(z)}{\sqrt{k}} + 2Q \phi(w)}
\left(-\frac12 \frac{\sqrt{k-2}}{(z-w)^2} +
\frac{\del \phi(w)}{z-w} \right)
\times \left( \alpha + i \sqrt{\frac{k-2}{k}} \right) \,,
\eea
and this fixes $\alpha = -i \sqrt{\frac{k-2}{k}}$. For this value of
$\alpha$, the OPE of $\psi^-$ with B is
\bea
\psi^-(z) B(w) \sim -\frac{i}{\sqrt{k}Q^2}
\del_{w} \left( \frac{e^{ 2Q \phi(w)} }{z-w}
\right) e^{-2i \frac{X(z)}{\sqrt{k}} }
\eea
so the integrated screening charge $\oint dw B(w)$ also commutes with
$\psi^-$. In the following we rescale $B$ by a constant and add the
antichiral factor, so we will use
\bea
B = \mu \left(i \sqrt{k} \del X_L + \frac{1}{Q}
\del \phi \right) \left( i \sqrt{k} \bar{\del} X_R +
\frac{1}{Q} \bar{\del} {\phi} \right)   e^{2 Q \phi}
\label{cigarinteraction}
\eea
%Another interaction which commutes with the parafermionic algebra
%was considered in \cite{Giribet:2001ft}, but we will not consider it here.
This, then, is the form of the black hole perturbation that is
consistent with the parafermionic symmetry. We will make use of this,
along with the Sine-Liouville operators of
Eqs.(\ref{sl1}),(\ref{sl2}), as screening charges in the following
subsections.

\subsection{Correlation functions}
Let us consider the  two-point function  of the interacting
$SL(2,R)/U(1)$ theory with  all the perturbations turned on.
The vertex operators $V_{j, m,\bar{m}}$ and $V_{-j+1, m,\bar{m}}$ have the same
conformal dimension and correspond to incoming and outgoing waves with
the same momentum.  We normalize them such that
\bea
\langle V_{-j+1, m,\bar{m}}  V_{j,-m,-\bar{m}} \rangle =1 \,,
\label{2pnorm}
\eea
and we will consider
the two-point function
\begin{eqnarray}
R(j,m,\bm) = \langle V_{j, m,\bar{m}}  V_{j,-m,-\bar{m}} \rangle \,.
\label{2pfunc}
\end{eqnarray}
We ignore the divergent delta functions in both (\ref{2pfunc}) and
(\ref{2pnorm}).  Using the $SL(2,R)$ quantum numbers is useful because
the coset theory inherits the structure of degenerate operators and
fusion rules of the $SL(2,R)$ algebra.  This in turn will allow us
compute (\ref{2pfunc}) by exploiting the trick of Teschner
\cite{Teschner:1995yf,Fateev:2000ik,Giveon:2001up}.

The affine $SL(2,R)$ algebra has degenerate primaries
at spins \cite{Kac:1979fz}
\bea
j_{r,s}  = -\frac{(r-1)}{2} - \frac{(s-1)}{2}k' \,,
\eea
where $k'\equiv k-2$ and $r,s$ are integers with either $r,s >0$ or $
r <0, s\leq 0$.  The OPE of a primary with spin $j_{r,s}$ gives only a
finite number of fields, according to fusion rules that were worked
out in
\cite{Awata:1992sm}.  Below we will consider the constraints on the
two-point function (\ref{2pfunc}) which follow from the degenerate
primaries with spins $j=-1/2$ and $j=-k'/2$.

\subsection*{The $j=-1/2$ degenerate field}
The fusion of the degenerate primary $V_{-\frac12,\frac12,\frac12}$
with any other primary gives \cite{Awata:1992sm}
\begin{eqnarray}
V_{-\frac12,\frac12,\frac12} V_{j,m,\bm} \sim
C^+_{j,m,\bm} \left[ V_{j-\frac12,m+\frac12,\bm+\frac12} \right]
+ C^{-}_{j,m,\bm}\left[V_{j+\frac12, m+\frac12,\bm+\frac12}\right]\,.
\label{fusion}
\end{eqnarray}
Consider the auxiliary three-point function
\begin{eqnarray}
\langle V_{j,m,\bm} (x_1), V_{j+\frac12,-m-\frac12,-\bm-\frac12}(x_2)
V_{-\frac12,\frac12,\frac12}(z) \rangle \,.
\end{eqnarray}
Taking $z\rightarrow x_1$ it is equal to
\begin{eqnarray}
C^-_{j,m,\bm} \, R\left(j+\frac12,m+\frac12,\bm+\frac12 \right)\,.
\end{eqnarray}
Taking $z\rightarrow x_2$ it is equal to
\begin{eqnarray}
C^+_{j,m,\bm} \, R(j,m,\bm)\,.
\end{eqnarray}
Equating the two expressions we get
\begin{eqnarray}
\frac{R(j+\frac12,m+\frac12,\bm+\frac12)}{R(j,m,\bm)}=
\frac{C^+_{j,m,\bm}}{C^-_{j,m,\bm}} \,.
\label{firstshifeq}
\end{eqnarray}
This is a functional equation for $R(j,m,\bm)$ that depends on the
structure constants $C^{\pm}_{j,m,\bm}$, which, from (\ref{2pnorm})
and (\ref{fusion}), are given by
\begin{eqnarray}
C^+_{j,m,\bm}&=&
\langle V_{-j+\frac{3}{2},-m-\frac12,-\bm -\frac12}(\infty)
V_{-\frac12,\frac12,\frac12}(1) V_{j,m,\bm}(0) \rangle \,, \\
C^{-}_{j,m,\bm} &=& \langle V_{-j+\frac12,-m-\frac12,-\bm-\frac12}
(\infty) V_{-\frac12,\frac12,\frac12}(1) V_{j,m,\bm}(0) \rangle \,,
\end{eqnarray}
where
\bea
V_{j,m,\bm} (\infty) = \lim_{z,\bar{z} \rightarrow \infty}
z^{2\Delta_{j,m}} \bar{z}^{2\bar{\Delta}_{j,\bm}} V_{j,m,\bm} (z,\bar{z})
\eea
is the standard BPZ conjugate. In this approach, the computation of
$C^{\pm}_{j,m,\bm}$ (and similar constants associated to the second
degenerate field below) is the only perturbative result needed, and
allows to compare the role of the black hole/Sine-Liouville
interactions.  The presence of the background charge $Q$ in the $\phi$
direction implies, for a correlator such as
\bea
\big\langle  \prod_{i=1}^n e^{2 \alpha_i Q\phi(z_i)}\big \rangle \,,
\eea
the anomalous conservation law
\bea
\sum_{i=1}^n \alpha_i = 1.
\label{anomalous}
\eea
{}From (\ref{primary}) it follows that (\ref{anomalous}) is satisfied
for $C^+_{j,m,\bm}$ without any insertion of the interactions, so we
have $C^+_{j,m,\bm}=1$.  For $C^-_{j,m,\bm}$, we can satisfy
(\ref{anomalous}) by inserting one cigar screening charge
(\ref{cigarinteraction}). This gives
\begin{eqnarray}
C^{-}_{j,m,\bm} &=&  \int d^2 z  \langle V_{-j+\frac12,-m-\frac12,-
\bm-\frac12}(\infty) B(z)
V_{\frac12,\frac12,\frac12}(1) V_{j,m,\bm}(0) \rangle_{free}
\eea
To compute the integrand, we use the free field correlators
\bea
 \langle e^{i\frac{2m+1}{\sqrt{k}}X_L(\infty)}
 e^{-\frac{i}{\sqrt{k}} X_L(1)}
e^{-\frac{2im}{\sqrt{k}}X_L(0)} \rangle &=& 1 \,,
\nn
\\
i \sqrt{k} \langle e^{i\frac{2m+1}{\sqrt{k}}X_L(\infty)}
\del X_L(z) e^{-\frac{i}{\sqrt{k}} X_L(1)}
e^{-\frac{2im}{\sqrt{k}}X_L(0)} \rangle &=&  -\frac{1/2}{z-1} - \frac{m}{z} \,,
\nn
\\
 \big \langle e^{(-2j+1)Q\phi(\infty)}
  e^{2Q\phi(z)}
e^{-Q\phi(1)}e^{2jQ\phi(0)}  \big \rangle &=&   z^{-2jQ^2}(z-1)^{Q^2} \,,
\nn
\\
\frac{1}{Q} \big \langle e^{(-2j+1)Q\phi(\infty)}
 \del \phi e^{2Q\phi(z)}
e^{-Q\phi(1)}e^{2jQ\phi(0)}  \big \rangle &=&  \left( \frac{1/2}{z-1}
- \frac{j}{z} \right)
\, z^{-2jQ^2}(z-1)^{Q^2},
\nn
\eea
and similar antiholomorphic expressions. This gives
\bea
C^{-}_{j,m,\bm} &=& \mu (m+j)(\bm+j) \int d^2z |z|^{-\frac{4j}{k-2}
-2} |z-1|^{\frac{2}{k-2}}
\nn \\
&=& - \mu \frac{\pi}{k'^2} (m+j)(\bm+j) \ga\left(-\frac{2j}{k'}\right)
\ga\left(\frac{2j-1}{k'}\right) \ga\left(1/k'\right) \,,
\label{cminus}
\end{eqnarray}
where $\ga(x) = \Gamma(x)/\Gamma(1-x)$ and we have used
(\ref{usefulintegral}).

We now show that one obtains the same expression for $C^{-}_{j,m,\bm}$
using the Sine-Liouville interactions of both dressings as screening
charges. The conservation law (\ref{anomalous}) can also be satisfied
by inserting one screening of type $S^1$ and one of type $S^2$, see
(\ref{sl1})-(\ref{sl2}).  This gives
\begin{eqnarray}
C^{-}_{j,m,\bm}    &=& \int d^2 z d^2 w
\langle V_{-j+\frac12,-m-\frac12,-\bm-\frac12}(\infty)
S^1_+ (w) S^2_-(z)
V_{-\frac12,\frac12,\frac12}(1) V_{j,m,\bm}(0) \rangle_{free} \nn \\
&=& \lambda_1 \lambda_2  \!\!
\int  \!\! d^2 z d^2 w  |z-w|^{-4} |z-1|^{\frac{2}{k-2}} \,
z^{m +j - \frac{2j}{k-2}} \, \bar{z}^{\bar{m} +j - \frac{2j}{k-2}}
\, w^{-m-j} \, \bar{w}^{-\bar{m}-j} \,\,.
\nn
\eea
To compute this integral we can
change variables from $(z,w)$ to $(z,y=w/z)$, and we get
\bea
C^{-}_{j,m,\bm} &=&
\lambda_1 \lambda_2  \!\!
\int  \!\! d^2 y |1-y|^{-4} y^{-m-j} \bar{y}^{-\bm -j} \times
\int  \!\! d^2 z |z|^{-2 -\frac{4j}{k-2}} |1-z|^{\frac{2}{k-2}} \,\,, \\
&=& - \lambda_1 \lambda_2 \,
\Ga(-1)\frac{\pi^2}{ k'^2}(m+j)(\bar{m}+j) \ga\left(-\frac{2j}{k'}\right)
\ga\left(\frac{2j-1}{k'}\right) \ga\left(1/k'\right) \,,
\label{cintegral}
\end{eqnarray}
%Using the free fields contractions
%\begin{eqnarray}
%\langle  e^{\alpha_1 \phi(z_1)} \ldots  e^{\alpha_n \phi(z_n)}\rangle
%= \Pi_{i<j} |z_i-z_j|^{-2\alpha_i \alpha_j}
%\end{eqnarray}
where we have used twice eq.(\ref{usefulintegral}). Thus we have
obtained precisely the same expression~(\ref{cminus}) for
$C^{-}_{j,m,\bm}$ using both Sine-Liouville screenings, and we can
identify
\bea
\mu = \lambda_1 \tilde{\lambda}_2 \,,
\label{mulambdas}
\eea
where
\bea
\tilde{\lambda}_2 = \pi \Gamma(-1)\lambda_2
\label{lambren}
\eea
is a renormalized value of $\lambda_2$. The reason to renormalize only
$\lambda_2$ in the product $\lambda_1 \lambda_2$ will become clear
below.

\subsection*{The $j=-k'/2$ degenerate field}
For this case, the fusion rules are \cite{Awata:1992sm}
\begin{eqnarray}
V_{-\frac{k'}{2},\frac{k'}{2},\frac{k'}{2}} V_{j,m,\bm} &\sim&
\tilde{C}^+_{j,m,\bm} \left[ V_{j-\frac{k'}{2},m
+\frac{k'}{2},\bm+\frac{k'}{2}} \right]
+ \tilde{C}^{-}_{j,m,\bm}\left[V_{j+\frac{k'}{2},
m+\frac{k'}{2},\bm+\frac{k'}{2}}\right]
\nn \\
&& \qquad + \, \tilde{C}^{\times}_{j,m,\bm}\left[V_{\frac{k'}{2}-j+1,
m+\frac{k'}{2},\bm+\frac{k'}{2}}\right] \,\,.
\end{eqnarray}
A similar reasoning as that used above leads  leads to the functional equation
\begin{eqnarray}
\frac{R(j+\frac{k'}{2},m+\frac{k'}{2},\bm+\frac{k'}{2})}{R(j,m,\bm)}
= \frac{\tilde{C}^+_{j,m,\bm}}{\tilde{C}^-_{j,m,\bm}} \,.
\label{secondshifeq}
\end{eqnarray}
As before, we can set
\bea
\tilde{C}^+_{j,m,\bm}&=&
\langle V_{-j+\frac{k'}{2}+1,-m-\frac{k'}{2},-\bm -\frac{k'}{2}}(\infty)
V_{-\frac{k'}{2},\frac{k'}{2},\frac{k'}{2}}(1) V_{j,m,\bm}(0) \rangle = 1 \,,
\eea
since the conservation law (\ref{anomalous}) is satisfied without any
perturbative insertion. As for
\bea
\tilde{C}^-_{j,m,\bm}&=&
\langle V_{-j-\frac{k'}{2}+1,-m-\frac{k'}{2},-\bm -\frac{k'}{2}}(\infty)
V_{-\frac{k'}{2},\frac{k'}{2},\frac{k'}{2}}(1) V_{j,m,\bm}(0) \rangle \,,
\eea
we can satisfy (\ref{anomalous}) by inserting two Sine-Liouville $S^1$
interactions.  This gives
\begin{eqnarray}
\tilde{C}^{-}_{j,m,\bm}    &=&
\int d^2 z d^2 w
\langle
V_{-j-\frac{k'}{2}+1,-m-\frac{k'}{2},-\bm -\frac{k'}{2}}(\infty)
S^1_- (w) S^1_+(z) V_{-\frac{k'}{2},\frac{k'}{2},\frac{k'}{2}}(1)
V_{j,m,\bm}(0) \rangle_{free} \nn \\
&=& \lambda_1^2   \!\! \int  \!\! d^2 z d^2 w  |z-w|^{-2k+2}
|z-1|^{2k'} \, z^{m -j}  \, \bar{z}^{\bar{m} -j }
\, w^{-m-j} \, \bar{w}^{-\bar{m}-j} \,\,.
\eea
Changing now variables from $(z,w)$ to $(z,y=w/z)$ we get
\bea
\tilde{C}^{-}_{j,m,\bm} &=&
\lambda_1^2 \int d^2 z |z|^{-4j-2k'} |z-1|^{2k'} \times
\int d^2 y^{-m-j} \bar{y}^{-\bm -j } |1-y|^{-2(k-1)} \,\,,
\nn \\
&=& \lambda_1^2 \pi^2  \ga\left(-2j-k'+1\right)
\ga\left(2j-1\right) \frac{\Ga(-m-j+1)}{\Ga(-m-j-k'+1)}
\frac{\Ga(\bm+j+k')}{\Ga(\bm+j)} \,,
\label{ctildeintegral}
\end{eqnarray}
where we have used (\ref{usefulintegral}) twice.
\vskip .4 cm
Now that we have the structure constants, the solution to
the functional equations (\ref{firstshifeq}) and (\ref{secondshifeq}) is
\begin{eqnarray}
R(j,m,\bm)
= (\mu \pi \ga(1/k'))^{1-2j}
\frac{\Gamma(2j-1)\Gamma(1+\frac{2j-1}{k-2})} {\Gamma(-2j+1)
\Gamma(1-\frac{2j-1}{k-2})}
\frac{\Gamma(-j+1+ \bm)\Gamma(-j+1-m)}{\Gamma(j+\bm)
\Gamma(j-m)} \nn \\
\label{2pf}
\end{eqnarray}
which is the expression we wrote above in (\ref{twopointf}).
Also, we get also the relation
\bea
\lambda_1^2 \pi^2 = (\mu \pi \ga(1/k'))^{k'}
\label{mulambdaone}
\eea
which was first obtained in \cite{Giveon:2001up} by similar methods.
Given this relation, it is clear that $\lambda_2$ rather than
$\lambda_1$ is the coefficient which should absorb the divergence coming
from $\Ga(-1)$ in~(\ref{cintegral}). But since from (\ref{mulambdas})
we see that $\tilde{\lambda}_2$ is finite, it follows from (\ref{lambren})
that $\lambda_2$ is effectively renormalized to zero, and therefore the second
Sine-Liouville screening disappears from the theory.

Also note from (\ref{mulambdas})
and (\ref{mulambdaone}) that only one of the three coefficients $\mu,
\lambda_1, \tilde{\lambda}_2$ is independent.  The expression
(\ref{2pf}) for $R(j,m,\bm)$ is symmetric under $m \leftrightarrow
\bm$ for $m-\bm \in \mathbb{Z}$, using (\ref{gamasin}). It satisfies
$R(-j+1,m,\bm) = R^{-1}(j,m,\bm)$, and for delta normalizable states
($j= \frac12 + i \mathbb{R}$) it is a phase, namely, the phase shift
between an incoming and an outgoing wave.

Using the Teschner trick one can also obtain the three-point function,
and the same special structure constants we computed above enter
similarly as an input for functional relations for the three-point
function, which follow from crossing symmetry of an auxiliary
four-point function \cite{Teschner:1997ft}.

Therefore, in this efficient approach to compute the correlators,
the role of the second Sine-Liouville dressing is established for two
and three point functions.  It would be interesting to use the second
Sine-Liouville screening to perform free-field computations similar to
those in \cite{Fukuda:2001jd}.

\section{Generalized FZZ Algebra}

As an application of the considerations detailed above, we will
investigate a generalized class of Sine-Liouville models. The FZZ
algebra procedure will then be employed to find the analogs of the
dual black hole operators. However, such operators exist only in the
special case of $c=1$ matter coupled to Liouville theory. Therefore we
will first give a brief survey of the relevant aspects of $c=1$ string
theory, including a listing of some interesting physical states
in the cohomology,  before proceeding to the model.

\subsection{Cohomology of $c=1$ strings}

The $c=1$ string is a special case of the linear dilaton background of
Section 2.2 where we set $Q=2$ to get a total central charge $c=26$.
With a cosmological perturbation to cut off the strong coupling
region, the worldsheet action is: \beq S_{\rm c=1}=\int
d^2z\left(-\del X{\bar\del}X + \del\phi {\bar\del}\phi + 2 {\hat
    R}(z,{\bar z})\phi + 4\pi\mu\,e^{2\phi}\right) \,.  \eeq The
string loop expansion in this theory is an expansion in
$\frac1{\mu^2}$.  The coordinate $X$ has the interpretation of time,
but in what follows we will consider its Euclidean continuation,
corresponding to the case of finite temperature. Thus $X$ is Euclidean
(spacelike) and compactified: \beq X(z,\zbar)\sim X(z,\zbar) + 2\pi R
\,.  \eeq The physical fields of the theory are defined by the BRST
procedure, which is most tractable when the worldsheet theory is a
free field theory. In the present case, the theory (at least in the
given variables) is free only at $\mu=0$, the limit in which the
effective string coupling is infinite. For this case, the BRST
cohomology has been worked out in
\cite{Mukherji:1991cp,Lian:1991ty,Bouwknegt:1991yg,%
  Polyakov:1991qx,Witten:1991zd,Witten:1992yj}. As observed in
\cite{Sen:2004yv}, at nonzero $\mu$ we can still use part of the
previous results.

Let us therefore start by reviewing the cohomology at $\mu=0$.  One
important class of physical operators\footnote{Here and in what
  follows, we refer to a dimension $(1,1)$ operator as a physical
  operator if it is BRST invariant after integration over the
  worldsheet. Typically such operators are also BRST invariant when
  multiplied by the ghost field combination $c\cbar$. We need to be
  more specific about the ghost dependence of a physical operator only
  if this dependence is nontrivial.} are the momentum ``tachyons'':
\beq T_\frac{n}{R}^\pm =
e^{i\frac{n}{R}X}\,e^{(2\mp\frac{n}{R})\phi},\quad n \in \mathbb{Z}
\eeq with left and right conformal dimensions equal to 1. These are
just the special cases of the operators already introduced in
Eqs.(\ref{slpertplus}),(\ref{slpertminus}).  As before, the superscripts
$\pm$ refer to non-normalizable/normalizable operators respectively.

Another important class of observables are the winding modes. Writing
$X=X_L+X_R$, we define $\tX = X_L-X_R$ in terms of which:
\beq
\cT_{nR}^\pm = e^{i\,nR\,\tX}\,e^{(2\mp nR)\,\phi},\quad n \in \mathbb{Z} \,.
\label{windingtach}
\eeq
These are clearly also $(1,1)$ operators.

The operators $T_{\frac{n}{R}}$ and $\cT_{nR}$ are dual to each other under
(timelike) T-duality:
\beq
X_R\to -X_R,\quad \phi\to\phi-\log R
\eeq
under which $X\to\tX$ and
\beq
R\rightarrow \frac{1}R, \quad \mu \rightarrow \mu R \,.
\eeq
Note that $T_0=\cT_0= e^{2\phi}$ is the cosmological operator.

There are other modes of dimension $(1,1)$.  They are called
``discrete states''\cite{Polyakov:1991qx,Witten:1991zd} and can be
thought of as two-dimensional ``remnants'' of the higher-spin fields
that exist in critical string theory. We start by writing the
following chiral operators at the self-dual radius $R=1$ \beq
W^\pm_{s,n}(z)= {\cal P}_{s,n}(\del^j X)\, e^{2in X_L}\,e^{(2\mp
  2s)\phi_L} \eeq where $s=0,\half,1,\ldots,$ and $n,n'=
s,s-1,\ldots,1-s,-s$ and ${\cal P}_{s,n}$ is a polynomial in
derivatives of $X_L$ with conformal dimension $s^2-n^2$. In
particular, ${\cal P}_{s,\pm s}=1$.

Because the above operators depend only on the left-moving part of the
Liouville field, which is a noncompact scalar field, they are not physical
operators. The physical operators are the combinations:
\beq
Y^\pm_{s;n,n'}(z,\zbar) = W^\pm_{s,n}(z)\Wbar^\pm_{s,n'}(\zbar) \,.
\eeq
For $n=n'=\pm s$ the above operators are the momentum modes $T_{\pm
2s}$, while for $n=-n'=\pm s$ they are the winding modes $\cT_{\pm
2s}$.  The remaining ones, with $n<s$ or $n'<s$ are the true discrete
states. The time-independent discrete states are those with
$n=n'=0$. Simple examples are the ones with $s=1,2$ for which the
relevant $c=1$ primaries are:
\bea
\label{expldisc}
\nn{\cal P}_{1,0} &=& \del X \,,\\
{\cal P}_{2,0} &=& (\del X)^4 + \sfrac32 (\del^2 X)^2 - 2 \del X \del^3 X \,,
\eea
and the corresponding non-normalizable discrete-state operators are:
\bea
\label{nonnormex}
Y^+_{1;0,0} &=& \del X {\bar\del} X\,, \\
Y^+_{2;0,0} &=& {\cal P}_{2,0}{\bar {\cal P}}_{2,0}\, e^{-2\phi}\,.
\eea
Although the above states have been tabulated for a specific radius
$R=1$, they will exist at other radii as long as $n+n'$ is an integer
multiple of $1/R$ and $n-n'$ is an integer multiple of $R$. In
particular this constraint is always satisfied for $n=n'=0$,
independent of the radius, hence the time-independent discrete states
$Y^+_{s;0,0}$ exist for all radius. Of course $s$ has to be an integer
in order for $n=n'=0$ to be allowed. Since we are working at general
values of the radius, we will concentrate on this set of
time-independent discrete states.

The first nontrivial state in this collection is just $Y^+_{1,0,0}=\del
X{\bar\del}X$, the radius-changing operator.  In the critical string
this would have just been the zero-momentum mode of the
graviton/dilaton. Here it is a ``remnant'' of those fields, and is
forced to have zero momentum. The other discrete states are similar
remnants of excited tensor states of the string, with fixed momenta.

Note that for the radius operator appearing in \eref{nonnormex} there
is a normalizable, or non-local, counterpart:
\beq
Y^-_{1,0,0}=\del X{\bar\del}X\,e^{4\phi} \,.
\eeq
This is precisely the black hole perturbation of the previous
sections, specialised to the case $Q=2$. It has been shown
\cite{Mukherji:1991kz} that starting from a perturbation of the $c=1$
string by $Y^-_{1,0,0}$, there is no obstruction to finding a
classical solution of closed string field theory (CSFT) to all orders
in $\alpha'$, and moreover the solution so obtained is
unique. Therefore, starting with
\eref{bhpert} one generates an exact (at tree level) CFT describing a
string background. It follows that this CFT must be the $SL(2,R)/U(1)$
black hole CFT. This closes the gap between the spacetime solution,
valid only for large $k$, and the CFT, which lacks a direct spacetime
interpretation.

But it also suggests a generalization. Observe that:
\beq
\label{gbhpert}
Y^-_{s;0,0} = {\cal P}_{s,0}{\bar{\cal P}}_{s,0} (\del^j X,
{\bar\del}^j X)\, e^{(2+2s)\phi} \eeq for $s=0,1,2,\ldots$ defines an
infinite family of normalizable operators, of which the first two
($s=0,1$) are the cosmological and black hole perturbations. Now the
considerations in Ref.\cite{Mukherji:1991kz} were shown to be
generally applicable to all these operators. Therefore each of them
similarly generates a unique classical solution of CSFT, and so
must correspond to some exact CFT. Unlike the first nontrivial
case ($s=1$, the usual 2d black hole) where the CFT is the
$SL(2,R)/U(1)$ nonlinear $\sigma$-model, the CFT in the other cases is
not explicitly known.  The form of the states in \eref{gbhpert}
suggests that we are dealing with higher-spin generalizations of the
2d black hole. As we will now argue, these are related by a
generalized FZZ duality to Sine-Liouville perturbations of higher
winding number.

\subsection{Higher Winding Sine-Liouville Perturbations}

Supposing that instead of the unit winding perturbation $V=\cT_1$, we
perturb the action of the linear dilaton theory by Sine-Liouville
operators of winding number 2:
\beq \cT^\pm_{\pm 2R} = e^{\pm
  2iR(X_L-X_R)}\,e^{(2\mp 2R)\phi}
\eeq
(recall that the $\pm$ sign in the subscript refers to the sign of the
winding number while the one in the superscript refers to the
dressing). It is easily checked that the OPE between mutually
conjugate operators of this type is again the black hole perturbation:
\beq
\cT^+_{2R}(z,\zbar)\cT^-_{-2R}(w,\wbar)\sim \frac{1}{|z-w|^2}
\del X{\bar\del}X\,e^{4\phi} + \cdots
\eeq
The same will be true for pairs of mutually conjugate operators of any
winding number -- in every case, the output of the OPE is the 2d black
hole perturbation.  One way to understand this is that we can orbifold
the compact time direction to enhance the radius by an integer factor.
The multiply wound Sine-Liouville perturbation of the original theory
then become singly-wound perturbations in the orbifolded theory. But
orbifolding in time does not affect the black hole perturbation
operator, which is time-independent.

Things become more interesting if we perturb the theory simultaneously
by operators of different winding numbers. As a first example,
consider the theory perturbed by the single and double-winding
operators: \beq
\label{multperturb}
S\to S + \int d^2z\, (\cT_R^+ + \cT_{-R}^+ + \cT_R^- + \cT_{-R}^- +
\cT_{2R}^+ + \cT_{-2R}^+ + \cT_{2R}^- + \cT_{-2R}^-)
\eeq
In this case, examining the OPE algebra, we find that the product of
{\it three} of these operators can potentially produce a new $(1,1)$
operator on the RHS:
\beq
\label{tripleope}
\cT_{-2R}^\pm(z_1,\zbar_1)\cT_{R}^\mp(z_2,\zbar_2)
\cT_{R}^\mp(z_3,\zbar_3)~\sim~
\cP_{2,0}(\del X){\bar\cP}_{2,0}({\bar\del}X)\,
e^{6\phi}=Y^-_{2,0,0}
\eeq
where $\cT_{nR}^\pm$ can be read off from \eref{windingtach} and
$\cP_{2,0}$ is given explicitly in \eref{expldisc}.

Let us work this out in more detail. We have:
\bea
\cT_{2R}^+(z_1,\zbar_1)\cT_{-R}^-(z_2,\zbar_2)
\cT_{-R}^-(z_3,\zbar_3) = \,\,:e^{2iRX_1}e^{(2-2R)\phi_1}:~
:e^{-iRX_2}e^{(2+R)\phi_2}:~
:e^{-iRX_3}e^{(2+R)\phi_3}:&&\nn\\
=\frac{1}{|z_{12}|^{4-2R}}\frac{1}{|z_{13}|^{4-2R}}\frac{1}{|z_{23}|^{4+4R}}\,
:e^{iR(2X_1-X_2-X_3)}:~:e^{(2-2R)\phi_1 +(2+R)\phi_2 + (2+R)\phi_3}:&&\nn\\
\eea
where we have used the shorthand $X_i\equiv X(z_i,\zbar_i)$ and
similarly for $\phi_i$, as well as $z_{ij}\equiv z_i -z_j$.

After integration over the $z_i$, the RHS of the above may be written
\beq
\int \prod_{i=1}^3 d^2 w_i \frac{1}{|w_1|^{4-2R}}\frac{1}{|w_2|^{4-2R}}
\frac{1}{|w_1-w_2|^{4+4R}} {\cal O}(w_i,\wbar_i)
\eeq
where we have defined $w_1\equiv z_1-z_2,w_2\equiv z_1-z_3,w_3=z_3$, and
\beq
\label{matterLiouville}
{\cal O}(w_i,\wbar_i)\equiv \left[
:e^{iR(2X(w_2+w_3)-
X(w_2-w_1+w_3)-X(w_3))}:\times~ (w\to\wbar)\right]:e^{6\phi(w_3,\wbar_3)}:
\eeq
Here we have moved all the Liouville fields $\phi_i$ to the location
$w_3$ and dropped the new terms that arise in doing this. As we will
see, at the end this will only lose us some terms that are trivial in
the BRS cohomology.

Finally we expand the $X$-dependent vertex operator about the point
$w_3$ as:
\beq
\left[:e^{iR(2X(w_2+w_3)-
X(w_2-w_1+w_3)-X(w_3))}:\times~ (w\to\wbar)\right]:
= \Big|\sum_{n_1,n_2=0}^\infty
w_2^{n_1} (w_2-w_1)^{n_2} {\cal A}_{n_1,n_2}(w_3)\Big|^2
\eeq
where the operators ${\cal A}_{n_1,n_2}$ are built out of holomorphic
derivatives of $X$, namely $\del X,\del^2 X,\cdots$ and have conformal
dimension $(\Delta,{\bar\Delta})=(n_1+n_2,0)$. Their complex
conjugates have dimension $(0,n_1+n_2)$. Anticipating that the final
contribution can only come from physical operators in the cohomology,
we keep only those operators in the sum which are of the form $|{\cal
A}_{n_1,n_2}|^2$ with $n_1+n_2=4$. Then the combined matter-Liouville
operator in
\eref{matterLiouville} can be replaced by:
\beq
|w_2|^{2n_1}|w_1-w_2|^{2n_2}|{\cal
A}_{n_1,n_2}(w_3)|^2~:e^{6\phi(w_3,\wbar_3)}:
\eeq
Now we see that the composite operator
\beq
|{\cal A}_{n_1,n_2}(w_3)|^2~:e^{6\phi(w_3,\wbar_3)}:
\eeq
has conformal dimension $(1,1)$ and is a local operator depending only
on $(w_3,\wbar_3)$. The coefficient functions depend only on $w_1,w_2$ and
combine under the integral sign into an expression of the form:
\beq
\int d^2 w_1 d^2 w_2 \frac{1}{|w_1|^\alpha}\frac{1}{|w_2|^\beta}
\frac{1}{|w_1-w_2|^\gamma}
\eeq
for some $\alpha,\beta,\gamma$ satisfying
$\alpha+\beta+\gamma=4$. Thus the coefficient of the $(1,1)$ operator
is logarithmically divergent, the sign of a nontrivial
$\beta$-function.

At this stage it is clear without further computation that the
operator ${\cal A}_{n_1,n_2}$ must be the Virasoro primary ${\cal
  P}_{2,0}$ defined in \eref{expldisc}. The reason is that the three
operators whose OPE we are computing are all in the cohomology and the
output must therefore also be in the cohomology. Given the total
matter and Liouville momenta of the fields on the LHS of the multiple
OPE, there is a unique such operator that can appear on the RHS. Hence
we have shown that the higher-spin black hole operator \beq
Y_{2,0,0}^-=\cP_{2,0}(\del X){\bar\cP}_{2,0}({\bar\del}X)\, e^{6\phi}
\eeq appears in the $\beta$-function of the theory perturbed as in
\eref{multperturb}, thereby justifying \eref{tripleope}.

The above result is quite general. For example, one can check that:
\beq
\cT_{NR}^\pm(z_1,\zbar_1)\cT_{-R}^\mp(z_2,\zbar_2)
\cdots\cT_{-R}^\mp(z_N,\zbar_N)~\sim~
\cP_{N,0}(\del X){\bar\cP}_{N,0}({\bar\del}X)\,e^{(2+2N)\phi}=Y^-_{N,0,0}
\eeq
Thus the higher-spin black hole operator of label $N$ (i.e. Liouville
momentum $2+2N$) arises when we perturb the linear dilaton theory with
Sine-Liouville operators of windings 1 and $N$.

We see that, in a similar sense as for the FZZ algebra of the previous
section, the multiply-wound Sine-Liouville operators are linked to
higher-spin black holes. More precisely, perturbing by all
Sine-Liouville operators of winding numbers $1,2,\cdots,N$ gives rise
to higher-spin black holes with all labels up to $N$ (the spins
realised in this way are $2k^2,~k=1,2\cdots,N$). This should be viewed
as a generalization of the FZZ duality,
and we expect that also here the coefficients of half of the Sine-Liouville operators
get renormalized to zero.

To produce only a definite
higher-spin black hole for $N\ge 2$, one must fine-tune the
perturbation strengths so that the lower-spin operators are not
produced.

\section{Conclusions}
In this work we presented a new approach to the FZZ duality between
the two-dimensional black hole and the sine-Liouville conformal field
theory. In this approach the duality is to be understood as coming
from the fact that the Sine-Liouville perturbations of both dressings
induce, via their mutual OPE, the operator representing a black hole
deformation, and one the of the two Sine-Liouville perturbations then disappears
because its coefficient gets renormalized to zero.

This approach has led us to propose a generalized FZZ duality for the $c=1$ string.
One side of this duality is a CFT generated by perturbing the linear
dilaton background with higher-spin analogues of the black hole
operator. Finding an exact description of this CFT would be very
helpful in understanding this generalized duality better, though that
appears to be a hard problem on which no progress has been made since
the existence proof in Ref.\cite{Mukherji:1991kz}. One
might instead try to use the holographic description in terms of
double-scaled matrix models to get more insight into the nature of the
theory that results from the fully back-reacted higher spin
perturbation. Also, even partial progress in the worldsheet treatment
of the higher spin perturbations (e.g. some correlation functions)
could provide relevant tests for the generalized FZZ duality we have
proposed.

\section*{Acknowledgements}

AP thanks Gast\'{o}n Giribet, Dan Isra\"{e}l, Martin Rocek and Jan
Troost for discussions on topics related to this paper. SM is grateful
to the Institute for Advanced Study, Princeton for hospitality while
part of this work was done. The research of AM is supported in part by
CSIR Award No. 9/9/256(SPM-5)/2K2/EMR-I. The research of AP is
supported by the Simons Foundation. We are grateful to the people of
India for generously supporting our research.

\appendix

\section{Useful formulae}

\begin{eqnarray}
\Ga(x)\Ga(1-x) &=& \frac{\pi}{\sin(\pi x)}
\label{gamasin} \\
\ga(x) &\equiv& \frac{\Gamma(x)}{\Gamma(1-x)} \\
%\ga(x)&=& \frac{1}{\ga(1-x)} \\
\ga(x+1) &=& -x^2 \ga(x) \\
%\end{eqnarray}
%
%\begin{eqnarray}
\int_{\mathbb{R}^2} \!\! d^2 x \,  x^{a} \bar{x}^{\bar{a}} (1-x)^{b}
(1-\bar{x})^{\bar{b}} &=&
\pi \frac{\Ga(1+a)}{\Ga(-\bar{a})} \frac{ \Ga(1+b)}{\Ga(-\bar{b})}
\frac{ \Ga(-\bar{a}-\bar{b}-1) }{  \Ga(a+b+2) }
\label{usefulintegral}\\
&=& (a,b \longleftrightarrow \bar{a},\bar{b})
\nn
\end{eqnarray}
The above integral is well defined only when $a-\bar{a}, b-\bar{b} \in
Z$, and it is only then that the second line holds.

\bibliographystyle{JHEP}

\bibliography{fzz}

\end{document}